# Effect of nonmagnetic substituents Mg and Zn on the phase competition in the multiferroic antiferromagnet MnWO$_4$

*Lynda Meddar[#], Michael Josse[§], Philippe Deniard[#], Carole La[#], Gilles André[+], Françoise Damay[+], Vaclav Petricek[@], Stéphane Jobic[#], Myung-Hwan Whangbo[&], Mario Maglione[§], and Christophe Payen[#,*]*

Institut des Matériaux Jean Rouxel (IMN), Université de Nantes-CNRS, Nantes, France, Institut de Chimie de la Matière Condensée de Bordeaux, CNRS, Université de Bordeaux 1, ICMCB, Bordeaux, France, Laboratoire Léon Brillouin, CEA-CNRS UMR 12, 91191 Gif-sur-Yvette cedex, France, Institute of Physics, Academy of Sciences of the Czech Republic, Praha, Department of Chemistry, North Carolina State University, Raleigh, North Carolina 27695-8204, USA

AUTHOR EMAIL ADDRESS. christophe.payen@cnrs-imn.fr

**RECEIVED DATE -**

TITLE RUNNING HEAD. Nonmagnetic substitution in multiferroic MnWO$_4$

CORRESPONDING AUTHOR FOOTNOTE.

[#] Institut des Matériaux Jean Rouxel (IMN)

[§] Institut de Chimie de la Matière Condensée de Bordeaux (ICMCB)

[+] Laboratoire Léon Brillouin (LLB)

[@] Academy of Sciences of the Czech Republic

[&] North Carolina State University

* Corresponding author. Email: christophe.payen@cnrs-imn.fr

ABSTRACT

The effects of substituting nonmagnetic $Mg^{2+}$ and $Zn^{2+}$ ions for the $Mn^{2+}$ (S = 5/2) ions on the structural, magnetic and dielectric properties of the multiferroic frustrated antiferromagnet $MnWO_4$ were investigated. Polycrystalline samples of $Mn_{1-x}Mg_xWO_4$ and $Mn_{1-x}Zn_xWO_4$ (0 ≤ x ≤ 0.3) solid solutions were prepared by a solid-state route and characterized via X-ray and neutron diffraction, magnetization, and dielectric permittivity measurements. Mg and Zn substitutions give rise to very similar effects. The Néel temperature $T_N$, the AF3-to-AF2 magnetic phase transition temperature $T_2$, and the critical ferroelectric temperature $T_c = T_2$ of $MnWO_4$ are reduced upon the nonmagnetic doping. At the lowest temperature (T = 1.5 K), the incommensurate magnetic structure for x(Mg) = 0.15 and x(Zn) = 0.15 corresponds to either a sinusoidal spin arrangement or an elliptical spin-spiral phase similar to the polar AF2 structure observed in $MnWO_4$. These findings were discussed by considering the effects of the Mg and Zn substitutions on the crystal lattice and on the spin exchange network of $MnWO_4$.

Nonmagnetic dilution effects explain the reduction of the magnetic and ferroelectric phase transition temperatures $T_N$ and $T_2$ by Mg or Zn substitutions at the Mn site in multiferroic $MnWO_4$

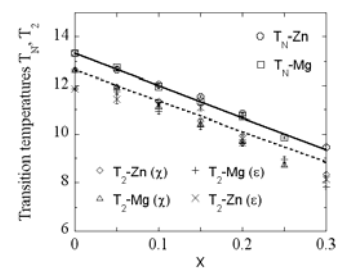

1. **Introduction**

Recently, the multiferroic materials, in which the ferroelectric polarization is driven by the magnetic ordering, have become a subject of much attention because of their magnetoelectic effect.[1] Many of these single-phase materials possess geometrically frustrated spin networks, which in general prevent the formation of conventional collinear spin structures. In this class of antiferromagnetic materials an incommensurate (ICM) magnetic structure with spiral-spin order can result, and the magnetic phase transition to this noncentrosymetrically ordered magnetic state can induce a spontaneous electric polarization via the spin-orbit coupling. The ferroelectric order is connected to the spin-spiral structure through a double vector product, $\mathbf{p} \propto \mathbf{e_{ij}} \times (\mathbf{S_i} \times \mathbf{S_j})$, where $\mathbf{p}$ is the local polarization and $\mathbf{e_{ij}}$ the unit vector connecting the nearest neighbor spins $\mathbf{S_i}$ and $\mathbf{S_j}$.[2,3]

Among the spin-spiral multiferroics, manganese tungstate $MnWO_4$, in which non-Jahn-Teller $Mn^{2+}$ ($d^5$) ions carry $S = 5/2$ spins, is an outstanding example.[4-6] Recent work has demonstrated that the ferroelectric and spin-spiral orders coexist and are intimately coupled in this material.[4-11] $MnWO_4$ undergoes three magnetic phase transitions in zero magnetic field below 14 K (Figure 1).[12] With decreasing temperature, $MnWO_4$ first transforms from a paramagnetic (PM) state to a collinear spin sinusoidal state (AF3) at $T_N \approx 13.5$ K, then to a tilted elliptical spiral spin state (AF2) at $T_2 \approx 12.3$ K, and eventually to a up-up-down-down collinear spin structure (AF1) at $T_1 \approx 8.0$ K. The magnetic structures of the AF3 and AF2 states are ICM to the lattice spacing with propagation vector $\mathbf{k} = (-0.214, 0.5, 0.457)$, while that of the AF1 state is commensurate (CM) with propagation vector $\mathbf{k} = (-0.25, 0.5, 0.5)$.[12] The loss of inversion symmetry due to the helical spin ordering at $T_2$ makes $MnWO_4$ exhibit ferroelectric polarization in the AF2 state, and $T_2$ is also the ferroelectric critical temperature.[4-6] In both the AF1 and AF3 states the Mn-spins have a collinear arrangement, and $MnWO_4$ does not show electric polarization since $\mathbf{p} \propto \mathbf{e_{ij}} \times (\mathbf{S_i} \times \mathbf{S_j}) = 0$ in these cases. Furthermore, $MnWO_4$ has a simple wolframite crystal structure, which is described by the monoclinic $P2/c$ space group.[13-14] The building blocks of $MnWO_4$ are $MnO_6$ octahedra containing $Mn^{2+}$ ions and $WO_6$ octahedra containing diamagnetic $W^{6+}$ ($d^0$) ions. The $MnO_6$ octahedra share edges to form zigzag $MnO_4$ chains along the c-direction (Figure 2a), and the $WO_6$ octahedra form zigzag $WO_4$ chains along the c-direction (Figure 2b). The three-dimensional (3D) structure of $MnWO_4$ is obtained from these $MnO_4$ and $WO_4$ chains on sharing their octahedral corners (Figure 2c). Thus, in $MnWO_4$, layers of magnetic $Mn^{2+}$ ions parallel to the bc-plane alternate with layers of diamagnetic $W^{6+}$ ions parallel to the bc-plane along the a-direction.

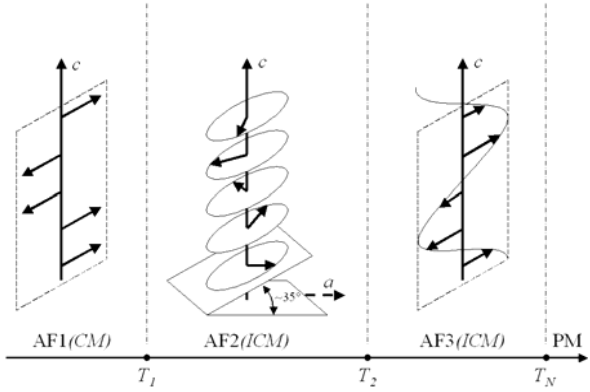

**Figure 1.** Schematic diagram showing the three ordered magnetic states of $MnWO_4$ below 14 K, where $T_N$ = 13.5 K, $T_2$ = 12.3 K, and $T_1$ = 8.0 K. In the collinear AF1 and AF3 magnetic states the Mn-spins are aligned with the easy axis of magnetization that lies in the ac plane at an angle of 35° with the a-axis. In the spin-spiral AF2 phase the Mn-spin has a additional component along b.

The existence of successive phase transitions in $MnWO_4$ has been attributed to competing magnetic interactions in the presence of weak magnetic anisotropy at each $Mn^{2+}$ site.[12,15] Indeed, the ICM components of the propagation vector of the AF3 and AF2 states, **k** = (-0.214, 0.5, 0.457), suggest that the spin interactions are frustrated along the a and c directions. It is well known that sufficiently strong frustration in a magnet results in a large number of quasi-degenerate low-energy states which can compete for the ground state. With decreasing temperature from the PM state, a weak single-ion anisotropy first selects the AF3 state in which Mn-spins are actually aligned along the easy axis of magnetization (Figure 1). This AF3 state is, however, "semi-ordered" as the spin fluctuations due to the frustration are still strong enough to induce a sinusoidal modulation of the Mn-spin amplitude. On further cooling, $MnWO_4$ enters the more ordered spin-spiral AF2 arrangement to reduce the extent of spin frustration. Magnetic entropy is further lowered on cooling by the onset of the phase transition to the AF1 state. This state has collinear spin order with no spin-amplitude reduction but is still spin-frustrated as it differs from the "up-down-up-down" Néel state. Besides, a recent DFT study revealed that the ICM state AF2 is more stable than the CM state AF1 in terms of the spin exchange interactions alone,[15] and the occurrence of the AF1 state below $T_1$ were attributed to the presence of weak magnetic anisotropy at each $Mn^{2+}$ site. Clearly, the phase competition in $MnWO_4$ should be modified by small perturbations like external pressure or intentional chemical doping. A substitutional doping should tune both the spin interactions and magnetic anisotropy and thus provides a way of better understanding the low-temperature properties of $MnWO_4$ and of modulating these properties. Actually, in $Mn_{1-x}Fe_xWO_4$ and $Mn_{1-x}Co_xWO_4$ solid solutions,[16,17] the substitution of magnetic $Fe^{2+}$ (S=2) or $Co^{2+}$ (S=3/2) ions for $Mn^{2+}$, which introduces extra Mn-M and M-M (M = Fe, Co) spin interactions along with a different local magnetic anisotropy determined by the $M^{2+}$ ions, modifies the phase competition. The Fe substitution stabilizes the CM AF1 order[16] whereas the Co doping suppresses this AF1 state and stabilizes the spin-spiral AF2 magnetic structure down to 4 K.[17] On the other hand, no study of the effect of nonmagnetic substitution on the multiferroïc properties of $MnWO_4$ has been reported so far.

In this paper, we report an initial study of the effect of nonmagnetic substitution at the Mn site on the structural, magnetic and dielectric properties of $MnWO_4$. The principal goal of this work was to test the robustness of the phase transitions and of the multiferroic state AF2 against nonmagnetic dopants. $Mg^{2+}$ and $Zn^{2+}$ ions were chosen because pure $MWO_4$ (M=Mg, Zn) and $MnWO_4$ are isostructural and posses only small differences in lattice cell volume of about 5 %. A moderate doping should therefore induce only little change in the crystal lattice of $MnWO_4$, and $MnWO_4$ should be diluted by Mg or Zn substituents without drastic changes of the magnetic interactions (e.g., due to local symmetry and bond-length variations). The use of two different dopants ($Mg^{2+}$ and $Zn^{2+}$) will help identify "universal" behaviors essentially due to the dilution of the magnetic network.

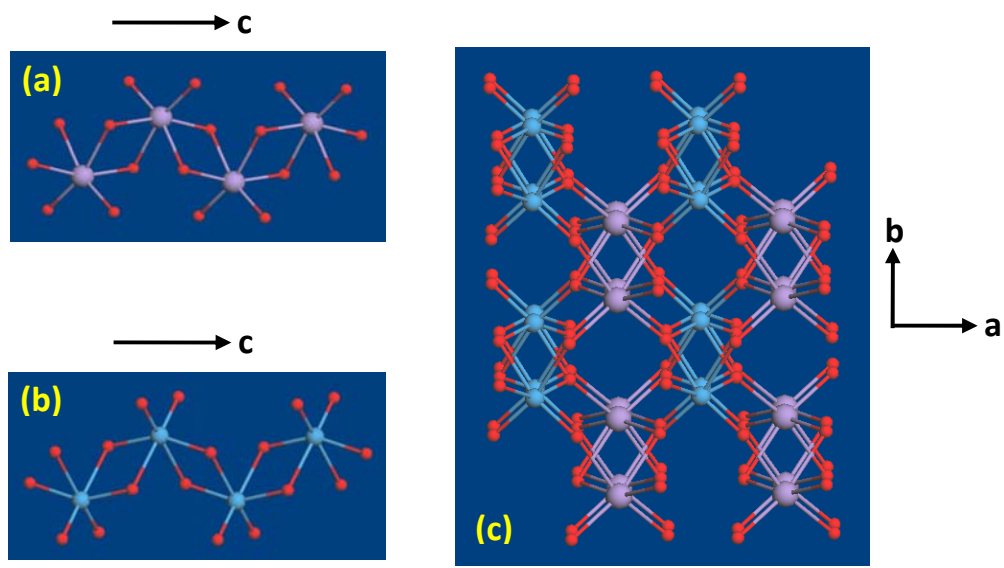

**Figure 2.** Perspective views of (a) a zigzag $MnO_4$ chain, (b) a zigzag $WO_4$ chain, and (c) the 3D arrangement of the $MnO_4$ and $WO_4$ chains in $MnWO_4$.

## 2. Experimental

Powder and ceramic samples of $Mn_{1-x}Mg_xWO_4$ and $Mn_{1-x}Zn_xWO_4$ (x = 0, 0.05, 0.10, 0.15, 0.20, 0.25, 0.30) solid solutions were prepared by a standard solid-state reaction method, starting from high purity MnO, $WO_3$, MgO or ZnO commercial powders with sub-micrometer particle size. The crystallinity and purity of the dried starting oxides were examined by X-ray diffraction (XRD) measurements. Stoichiometric amounts of these oxides were ball milled in ethanol for several hours, dried, and then pressed to form pellets. The pellets were heated at 850 °C for 30 h in air with an intermediate mechanical grinding. The pellets were then ground and the resultant powders were used to make dense ceramics suitable for dielectric permittivity measurements. These disk-shaped samples were sintered at temperatures between 1000 and 1150 °C for 1h in air. The sintering behavior of the samples was studied by measuring the change in length of a cylindrical sample with increasing temperature using a commercial dilatometer. The relative density of all sintered samples was higher than 90%.

Chemical analyses were performed using energy dispersive X-ray spectroscopy at different positions on the sample surfaces. Within the experimental accuracy of a few percent, the results agree with the nominal concentrations of the metal atoms (i.e., Mn, W, Mg or Zn). For the structural analysis, powder XRD data were collected at room temperature with a 2θ scan step of 0.0084° on a Bruker diffractometer (D8 Advance) equipped with a Vantec position sensitive detector using Cu $K_{\alpha 1}$ radiation (germanium monochromator, λ = 1.540598 Å). Neutron powder diffraction (NPD) patterns of 6 g samples of $Mn_{1-x}Zn_xWO_4$ and $Mn_{1-x}Mg_xWO_4$ with nominal x = 0.15 were taken at LLB. High-resolution diffraction data were collected at T = 300 K on the 3T2 diffractometer (incoming wavelength λ = 1.22537 Å). The magnetic structures were examined using the G4.1 instrument (incoming wavelength λ = 2.4226 Å). XRD and NPD patterns were analyzed using JANA 2006 and FULLPROF programs.[18,19]

A SQUID magnetometer (MPMS Quantum Design) was used to investigate the magnetic properties of the specimens. The temperature dependence of the zero-field-cooled (ZFC) and field-cooled (FC) dc magnetization was measured down to 2 K. In the FC mode, the applied field is switched on in the paramagnetic regime and the measurements are made while cooling across the transition temperatures to 2 K. The ZFC heating and FC cooling rates were ±0.03 K/min. The susceptibility was defined as the ratio of the dc magnetization *M* to the applied field *H* (i.e., χ=*M*/*H*).

Dielectric measurements were performed at ICMCB on sintered discs using an HP4194a impedance bridge. Samples were placed in a Quantum Design Physical Properties Measurement System (PPMS). Prior to these measurements, after deposing gold electrodes on the circular faces of the disks by cathodic sputtering, silver wires were attached on the top and the bottom of the pellets using a silver paste. These measurements were carried out in the frequency range of $10^2$ -$10^3$ kHz and in the temperature range of 4-16 K. All the capacitances and loss tan δ data were collected at heating and cooling rates of 0.2 K/min and at zero magnetic field. All samples displayed very small capacitances (~7 pF), and the ferroelectric transition was associated with even smaller variations of the capacitance (typically a few fF).

### 3. Results

**3.1 Room temperature X-ray and neutron diffraction.** Figure 3 presents the room-temperature XRD patterns of the $Mn_{1-x}Mg_xWO_4$ and $Mn_{1-x}Zn_xWO_4$ powder samples (x ≤ 0.3). All patterns show very narrow diffraction peaks without any splitting or extra reflection, and are consistent with those reported for the monoclinic *P2/c* Wolframite structure of $MnWO_4$ (JCPDS 13-0434). Rietveld refinements of these patterns were performed on the basis of the literature *P2/c* structural model of $MnWO_4$ using a random distribution of the Mn and Mg or Zn ions on the Wyckoff 2f position. All positions were fully occupied. Due to the presence of the heavy $W^{6+}$ ion and the small difference in the numbers of electrons between $Mn^{2+}$ and $Zn^{2+}$ ions, the fractional occupancies for Mg or Zn at the 2f Mn site were held fixed to the nominal Mg or Zn molar content x. With this model, all patterns could be successfully refined. The M-O (M = Mn, Mg, Zn) and W-O bond lengths calculated from the refined lattice parameters and atomic coordinates are in good agreement with those observed for Wolframite $MWO_4$ (M = Mn, Mg,

Zn). Furthermore, the corresponding bond valence sum calculations[20] are consistent with the presence of $M^{2+}$ (M = Mn, Mg, Zn), $W^{6+}$, and $O^{2-}$ ions.

Figure 4 shows the refined lattice parameters and the calculated cell volume of $Mn_{1-x}M_xWO_4$ (M = Zn or Mg) as a function of the nominal molar concentration x. As expected from Vegard's law, the lattice parameters ($a$, $b$, and $c$) decrease with increasing x, because $Mg^{2+}$ or $Zn^{2+}$ ions are smaller in size than $Mn^{2+}$; the ionic radii of $Mg^{2+}$ and $Zn^{2+}$ ions at an octahedral site are 0.72 and 0.74 Å, respectively, while that of a high-spin $Mn^{2+}$ ion at an octahedral site is 0.83 Å.[21] Furthermore, the cell volume of $Mn_{1-x}M_xWO_4$ (M = Zn or Mg) evolves linearly as the weighted average between those of $MnWO_4$ and $MWO_4$ (M = Zn or Mg), $V(x)=(1-x)*V(MnWO_4)+x*V(MWO_4)$, suggesting that the actual concentration of M in $Mn_{1-x}M_xWO_4$ (M = Zn or Mg) is equal, or very close, to the nominal value x.

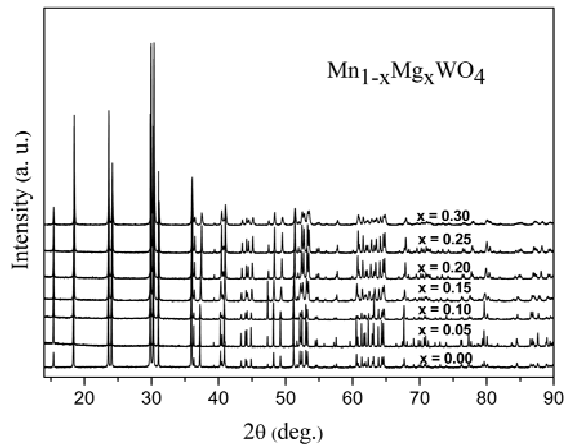

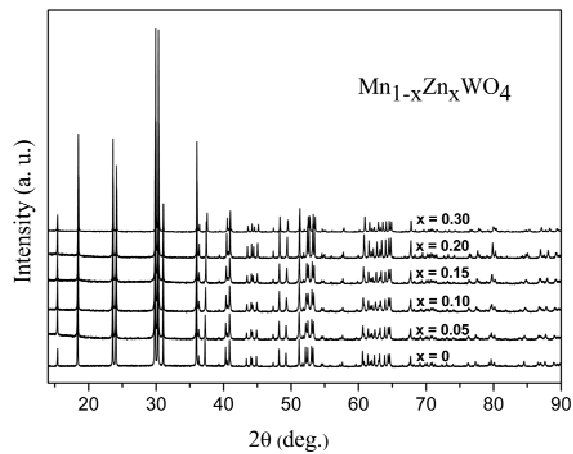

**Figure 3.** Room-temperature XRD patterns of $Mn_{1-x}Mg_xWO_4$ and $Mn_{1-x}Zn_xWO_4$ powder samples (x ≤ 0.3).

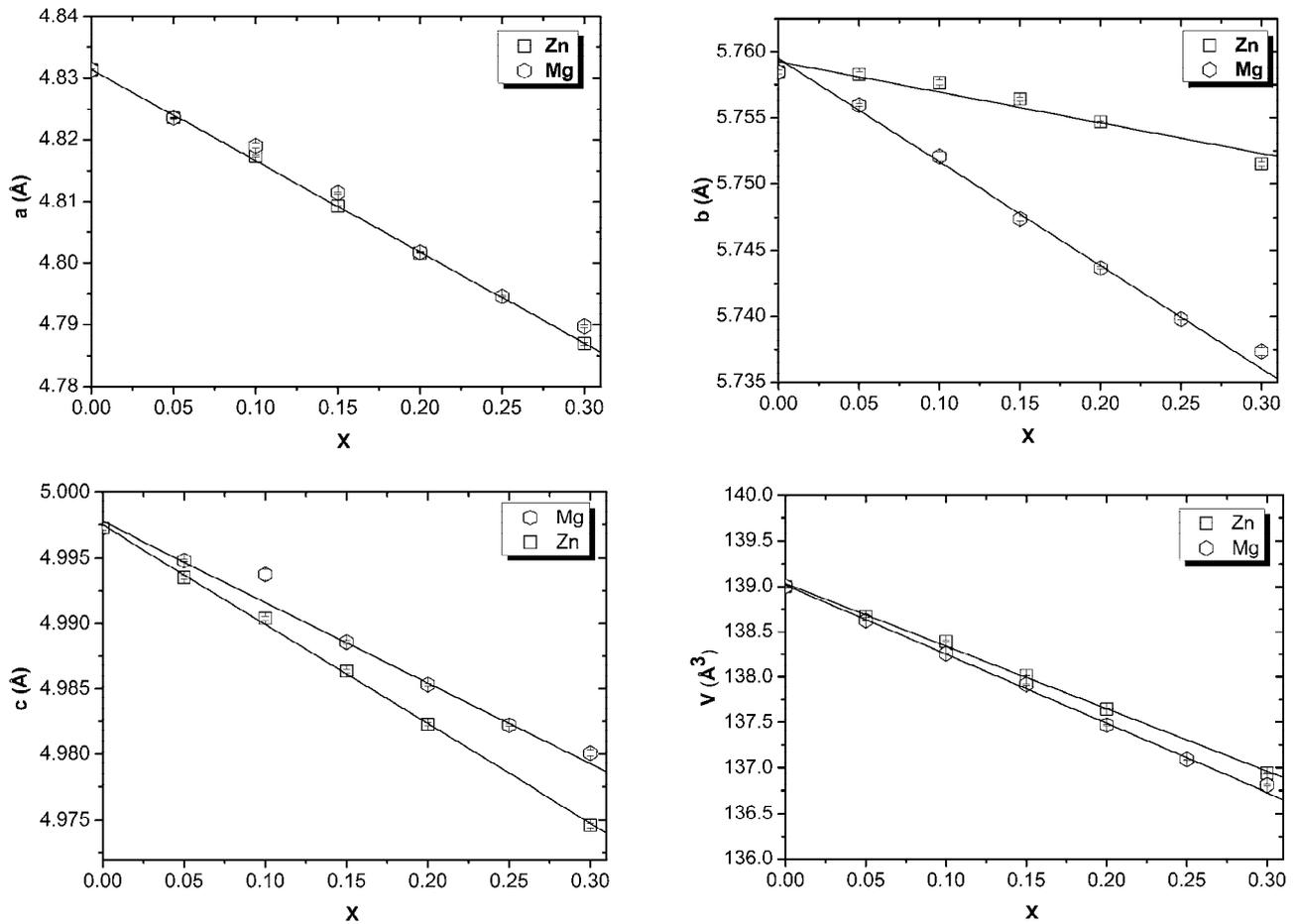

**Figure 4.** Refined lattice parameters and cell volumes of $Mn_{1-x}M_xWO_4$ (M = Mg, Zn) powder samples (x ≤ 0.3) as a function of $x$. The solid lines are drawn through the lattice parameter points as a guide for the eyes. The solid lines drawn in the cell-volume plots refer to the relationship, $V(x) = (1-x)*V(MnWO_4) + x*V(MWO_4)$, for M = Mg and Zn. The cell volumes of the pure $MWO_4$ (M = Mg, Zn) phases were taken from the literature.[22]

Since the Mg or Zn content could not be determined from the refinements of the XRD patterns, NPD data of a sample of $Mn_{1-x}Zn_xWO_4$ with nominal x = 0.15 were collected at room temperature (Figure 5). Rietveld refinements were carried out on the basis of the results obtained from the XRD measurements using the coherent nuclear scattering lengths b(Mn) = -3.73 fm, b(Zn) = 5.68 fm, b(W) = 4.86 fm, and b(O) = 5.803 fm. It is clear that the precision in the determination of the Mn/Zn occupation at the 2f site should be satisfactory due to the contrast in scattering lengths between Mn and Zn. With the occupancies of the W and O sites kept fixed, the Rietveld agreements factors were $R_p$ = 1.84, $R_{wp}$ = 2.38, GOF = 0.67. The refined lattice parameters, atomic positions, bond distances and bond valence sums are similar to those obtained from XRD refinement. The refined Zn occupancy at the 2f site, 0.149(4), is in excellent agreement with the nominal value x = 0.15. Attempts to refine the NPD pattern with W atoms at the Mn 2f site and with Zn or Mn atoms at the W atom sites yielded worse results. Hereafter, therefore, we will consider that the actual Mg or Zn concentrations are equal to their nominal values x.

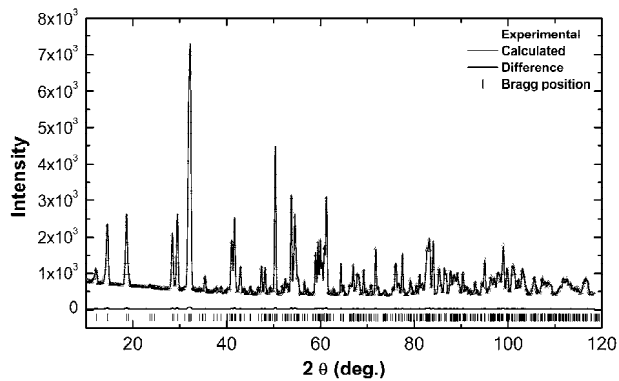

**Figure 5.** Observed (triangles) and calculated (solid curve) neutron powder diffraction patterns of $Mn_{0.85}Zn_{0.15}WO_4$ collected on 3T2 of the LLB at 300 K using neutrons of λ = 1.2254 Å. The tick-marks indicate the positions of the nuclear Bragg reflections. The lower curve shows the difference between the observed and calculated data on the same scale.

**3.2 Bulk magnetic properties.** We first examine the bulk magnetic properties at temperatures above the magnetic phase-transition temperatures of MnWO$_4$ (T$_N$ = 13.5 K). Figure 6 shows the reciprocal susceptibility data of powder samples of Mn$_{1-x}$Mg$_x$WO$_4$ and Mn$_{1-x}$Zn$_x$WO$_4$ (x ≤ 0.3). Above T ≈ 20 K, all samples exhibit a Curie-Weiss behavior, $1/\chi(x,T) = [T-\theta(x)]/C(x)$. The molar Curie constants $C(x)$ and the absolute Weiss temperatures $|\theta(x)|$ obtained from Curie-Weiss fits to the high temperature susceptibility data (T > 150 K) are plotted as a function of x in Figure 7. The results for MnWO$_4$, namely, $C(0) \approx 4.25$ cm$^3$ K mol$^{-1}$ and $\theta(0) \approx -71$ K, compare well with those reported in the literature.[4,23] For both nonmagnetic dopants, the molar Curie constant $C(x)$ and the absolute Weiss temperature $|\theta(x)|$ decrease linearly with increasing x, as $C(x) = (1-x)*C(0)$ and $|\theta(x)| = (1-x)*|\theta(0)|$, respectively (Figure 7). These linear behaviors are expected in the high-temperature limit for any randomly diluted antiferromagnet.[24] As a matter of fact, the Curie-Weiss fits could be done above T ≈ 2*$|\theta(x)|$ where the Curie-Weiss law is strictly applicable. In this mean-field regime, the Weiss temperature is given by the well-known relationship $\theta = \frac{S(S+1)}{3k_B}\sum_i z_i J_i$, where $J_i$ is the exchange coupling between a central spin and the $z_i$ spins linked by $J_i$. Within this approach, the dilution-induced decrease in $|\theta(x)|$ can be explained solely in terms of the increase in the missing magnetic bonds caused by the nonmagnetic substitution. The observed Weiss temperatures are negative, showing that the principal magnetic interactions remain antiferromagnetic in the diluted systems. The effective moments per Mn$^{2+}$ calculated from the Curie constant $C(x)$ are in the range of 5.75 - 5.9 $\mu_B$, which is consistent with the spin only value of 5.92 $\mu_B$.

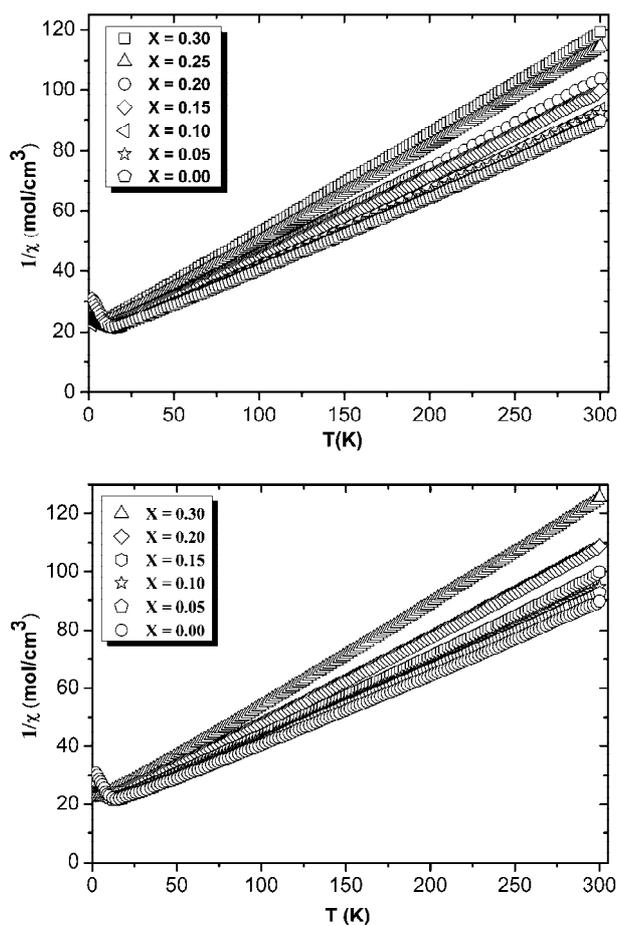

**Figure 6.** Temperature dependence of the reciprocal ZFC magnetic susceptibility $1/\chi(x,T)$ determined for the powder samples of $Mn_{1-x}Mg_xWO_4$ (top) and $Mn_{1-x}Zn_xWO_4$ (bottom) ($x \leq 0.3$) obtained at $\mu_0 H = 0.1$ T.

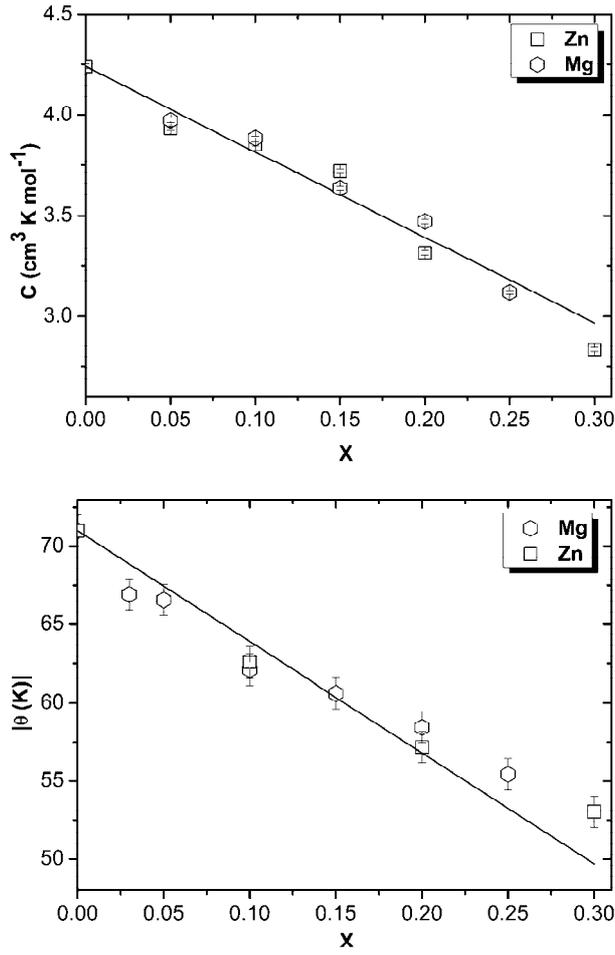

**Figure 7.** Molar Curie constants $C(x)$ (top) and absolute Weiss temperatures $|\theta(x)|$ (bottom) of $Mn_{1-x}M_xWO_4$ (M = Mg, Zn) as a function of x. The solid line refers to the relationship $C(x) = (1-x)*C(0)$ for the Curie constants, and $|\theta(x)| = (1-x)*|\theta(0)|$ for the Weiss temperatures.

We now analyze the low-temperature magnetic susceptibility data of $Mn_{1-x}Mg_xWO_4$ and $Mn_{1-x}Zn_xWO_4$ powder samples ($0 \leq x \leq 0.3$), shown in Figures 8 and 9. The data for powder $MnWO_4$ (x = 0) sample are consistent with those obtained for single-crystal samples,[4,5] showing three magnetic phase transitions at $T_1 \approx 7.5$ K, $T_2 \approx 12.5$ K, $T_N \approx 13.5$ K, which we estimated from the peak positions in the derivative $d\chi/dT$ curve obtained from the ZFC heating curve (Figure 8). A small but significant thermal hysteresis around $T_1$ is evidenced by comparing the ZFC heating and the FC cooling curves (Figure 8). This feature is fully consistent with the first-order nature of the AF1 to AF2 phase-transition in MnWO$_4$. For the substituted compounds, no difference between the ZFC heating and FC cooling traces was observed down to 2 K, for all x values up to the highest value of 0.30. As shown in Figure 9, the substituted samples do not show any anomaly in the susceptibility that can correspond to the AF1-to-AF2 phase transition observed for MnWO$_4$. This is also clearly seen in the derivative d$\chi$/dT curves shown in Figure 10. For both nonmagnetic ions Zn$^{2+}$ and Mg$^{2+}$, both the AF2-to-AF3 transition temperature $T_2$ and the Néel temperature $T_N$ decrease with increasing x both at the rate of about –0.13 K per 1% mol of Mg or Zn. The values of $T_2$ and $T_N$ are plotted as a function of x in Figure 11. Except for x(Mg) = 0.30, the peak anomalies for $T_2$ and $T_N$ are well defined hence leading to accurate determinations of these transition temperatures. This also indicates that the prepared samples are homogeneous. Overall, the magnetic properties of the substituted compounds appear to be independent of the nature of the nonmagnetic dopant.

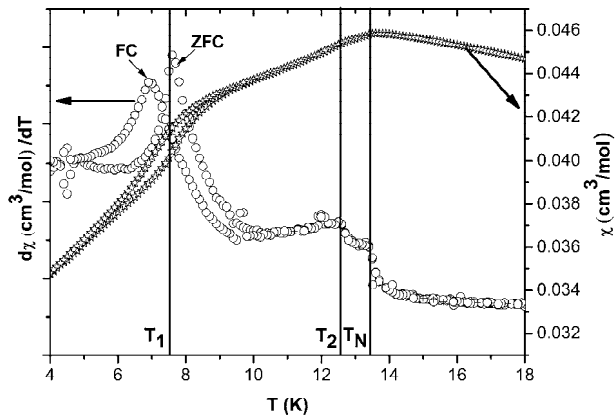

**Figure 8.** The temperature dependence of the ZFC and FC magnetic susceptibility of a powder sample of MnWO$_4$ obtained at $\mu_0H$ = 0.1 T (star symbols). Open circles show the corresponding derivative d$\chi$/dT curves. The vertical lines indicate the magnetic phase-transition temperatures as determined from the peaks in the ZFC derivative.

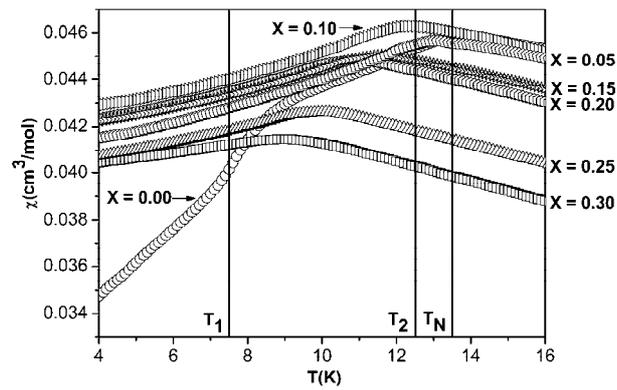

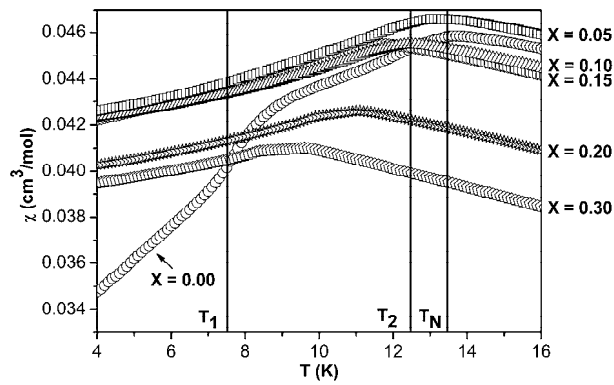

**Figure 9.** Temperature dependence of the ZFC molar magnetic susceptibility of $Mn_{1-x}Mg_xWO_4$ (top) and $Mn_{1-x}Zn_xWO_4$ (bottom) powder samples (x ≤ 0.3) obtained at $\mu_0H = 0.1$ T. The vertical lines indicate the magnetic phase-transition temperatures for $MnWO_4$.

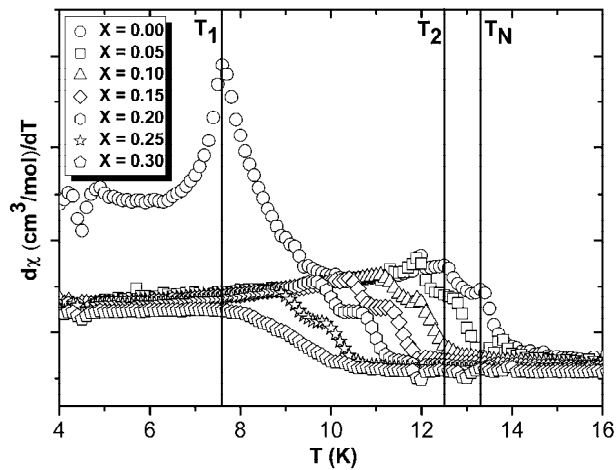

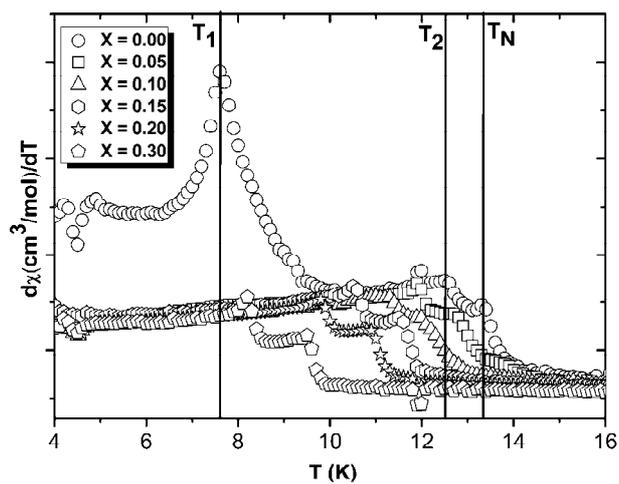

**Figure 10.** Derivative of the temperature-dependent ZFC magnetic susceptibility of $Mn_{1-x}Mg_xWO_4$ (top) and $Mn_{1-x}Zn_xWO_4$ (bottom) powder samples (x ≤ 0.3). The vertical lines indicate the magnetic phase-transition temperatures of $MnWO_4$.

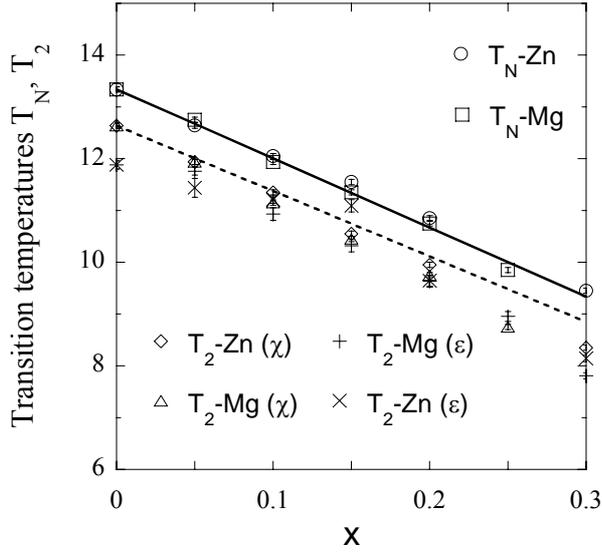

**Figure 11.** AF2-to-AF3 and Néel transition temperatures ($T_2$ and $T_N$, respectively) plotted as a function of Mg or Zn concentration x. $T_2(\chi)$ and $T_2(\varepsilon)$ are the values determined from the magnetic susceptibility and dielectric measurements, respectively. The prediction of mean-field theory for $T_N$ and $T_2(\chi)$ reductions of randomly diluted magnets are given by the solid and dashed lines, respectively. For the description of $T_2(\varepsilon)$, see Section 3.3.

**3.3 Dielectric properties at zero magnetic field.** In order to gain insight into ferroelectric transitions and possible dielectric dispersions at low temperature, the dielectric responses of dense ceramic samples of $Mn_{1-x}M_xWO_4$ (M = Mg, Zn) were investigated between 100 kHz and 1 MHz at temperatures from 4 to 16 K and at zero magnetic field. No dielectric dispersion was observed in these samples, whatever the experimental conditions. This lack of dispersion is a proof of the ferroelectric nature of the observed dielectric anomalies. For a given frequency, tiny differences between the heating and cooling data were observed. These differences corresponded to a thermal hysteresis of about 0.05 K and were due to the value of the heating and cooling rate of 0.2 K/min rather than a physical hysteresis. The temperature-dependent capacitance and loss tangent of the $Mn_{1-x}Mg_xWO_4$ and $Mn_{1-x}Zn_xWO_4$ ceramics (x ≤ 0.3) are shown in Figures 12 and 13, respectively. For $MnWO_4$, a sharp peak is observed in both the capacitance and loss tangent at $T_2 \approx 12$ K, consistent with the results obtained on single crystals.[5] With decreasing temperature, there is a decrease in the capacitance that corresponds to the stepwise transition previously seen at $T_1 \approx 7.5$ K on single crystals, whatever the crystal orientation.[5] This transition is due to the disappearance of the ferroelectric polarization when cooling across the AF2-to-AF1 transition. Here the broadening of this anomaly can be related to the polycrystalline state. A small step is nonetheless observed in the loss factor at $T_1 \approx 7.9$ K (Figure 14).

For all the substituted samples, the sharp peak associated with the AF3-to-AF2 transition preserves its shape upon doping. The peak position in the capacitance is the same as that in the dissipation factor, and both peaks shift gradually to lower temperature with increasing substitution. The widths of the peaks do not change significantly with x (except for the case of x = 0.05, which is reproducible but unaccounted for). The peak positions are independent of frequency. Thus these maxima in the capacitance and loss tangent correspond to the ferroelectric transition temperatures $T_2(\varepsilon)$. As shown in Figure 11, the x-dependence of $T_2(\varepsilon)$, determined from the dielectric properties, is practically the same as that of $T_2(\chi)$, determined from magnetic susceptibilities. In the Mg-substituted samples with $x \leq 0.15$, an additional low-temperature anomaly is detected below $T_2(\varepsilon)$, either as a small peak in capacitance (x(Mg) = 0.10 and 0.15) or a weak stepwise anomaly in loss factor (x(Mg) = 0.05), at $T_{Mg}(\varepsilon) \approx 7.6$, 6.7, and 6.1 K for x(Mg) = 0.05, 0.10, and 0.15, respectively (Figure 14). It is tempting to attribute this anomaly to the AF2-to-AF1 transition observed at $T_1 \approx 7.9$ K in $MnWO_4$. No corresponding anomaly in the magnetic response has been seen from bulk magnetization measurements and neutron diffraction (see Section 3.4), but considering the small amplitude of capacitance peak in x(Mg) = 0.10 and x(Mg) = 0.15 samples (below 1fF), these techniques may not be sensitive enough to detect it, if there is one. For the Zn-doped samples, no similar peak or stepwise anomaly is observed, either in the capacitance or in the loss factor.

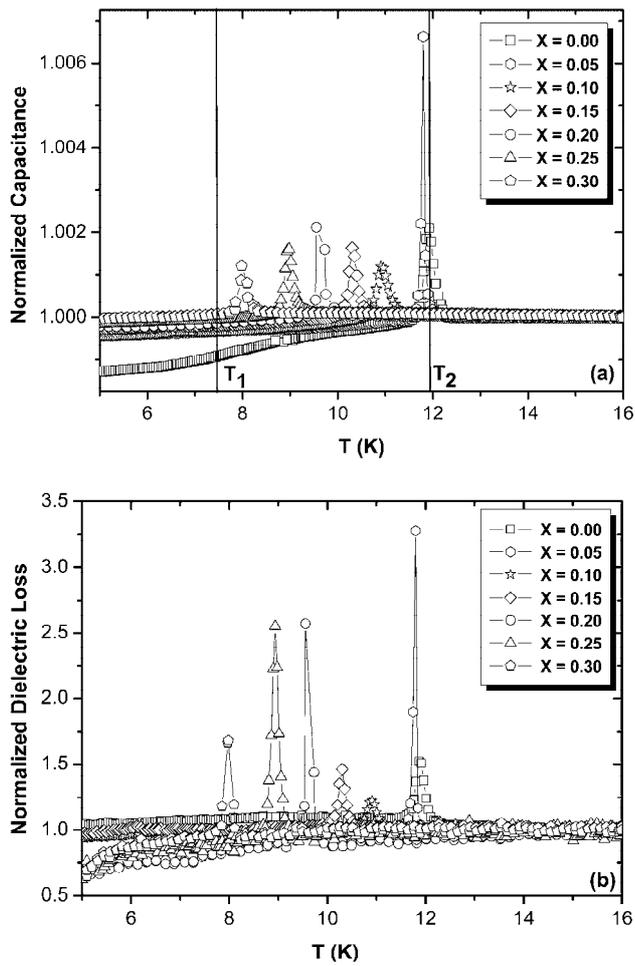

**Figure 12.** Temperature dependence of the capacitance and the loss factor of $Mn_{1-x}Mg_xWO_4$ ceramic samples (x ≤ 0.3), measured at 788 kHz during the heating run. The capacitance and loss tangent data were arbitrarily normalized to the values measured at 16 K.

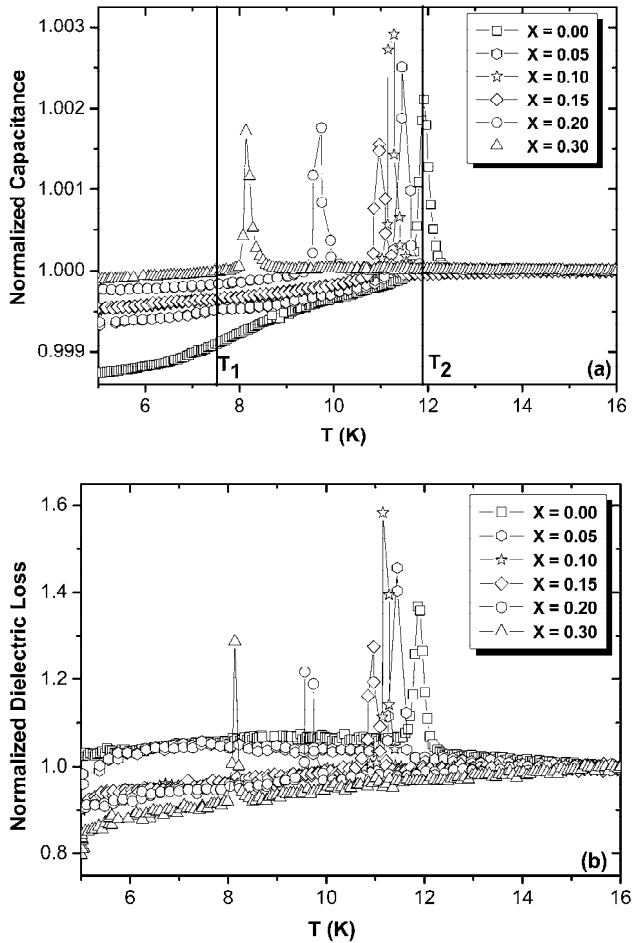

**Figure 13.** Temperature dependence of the capacitance and the loss factor of $Mn_{1-x}Zn_xWO_4$ ceramic samples ($x \leq 0.3$), measured at 788 kHz during the heating run. The capacitance and loss tangent data were arbitrarily normalized to the values measured at 16 K.

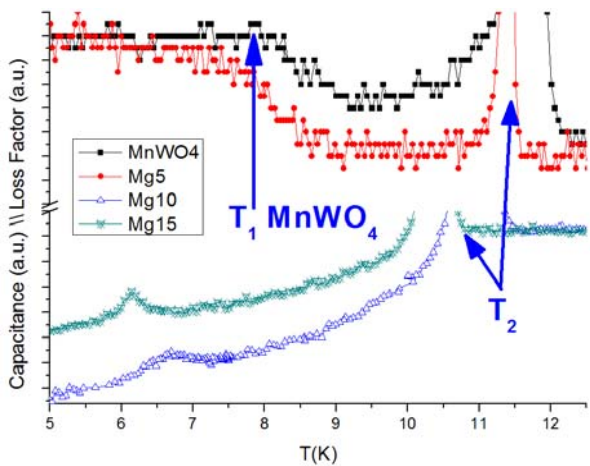

**Figure 14.** Low-temperature anomaly in the capacitance or loss factor for $Mn_{1-x}Mg_xWO_4$ ($x$ = 0.05, 0.10, 0.15) ceramic samples

**3.4 Low-temperature powder neutron diffraction.** Figure 15 shows the thermal evolution of the NPD pattern of $Mn_{0.85}Zn_{0.15}WO_4$ in the temperature range of 1.5 – 15 K. These data were taken by using the sample that was previously examined by high-resolution neutron diffraction at room temperature (see Section 3.1). The relative intensities of the magnetic peaks in the NPD patterns collected for $Mn_{0.85}Mg_{0.15}WO_4$ in the same temperature range of 1.5 – 15 K are identical to those observed for $Mn_{0.85}Zn_{0.15}WO_4$. For both substituted samples, the AF1 phase does not show up in the NPD above 1.5 K. Above the macroscopic Néel temperature $T_N(x = 0.15) \approx 12$ K, the patterns show only the peaks expected from the room-temperature crystal structure. In addition, a diffuse scattering with a form similar to that observed in $MnWO_4$[12] is present over a notable 2θ range centred approximately at 18°. This 2θ corresponds to an elastic wave vector $Q_{el} \approx 0.8$ Å$^{-1}$, which is roughly the same as in $MnWO_4$. As in $MnWO_4$, this diffuse scattering disappears on cooling below about 10 K. Therefore the diffuse scattering indicates the presence of short-range antiferromagnetic correlations that develop above the Néel temperature and disappear only below $T_2$ in $MnWO_4$ or $T_2(x = 0.15) \approx 10.6$ K in $Mn_{0.85}M_{0.15}WO_4$ (M = Mg, Zn). This disappearance corresponds to the onset of long range spin-spin correlations which do not exist even in the semi-ordered sinusoidal AF3 state. Between $T_N(x = 0.15)$ and $T_2(x = 0.15) \approx 10.6$ K, the NPD patterns contain several new magnetic Bragg peaks, consistent with the incommensurate AF3 magnetic structure with **k** = (-0.214, 0.5, 0.457) observed for $MnWO_4$ between $T_N$ and $T_2$ (Figure 15).[12]

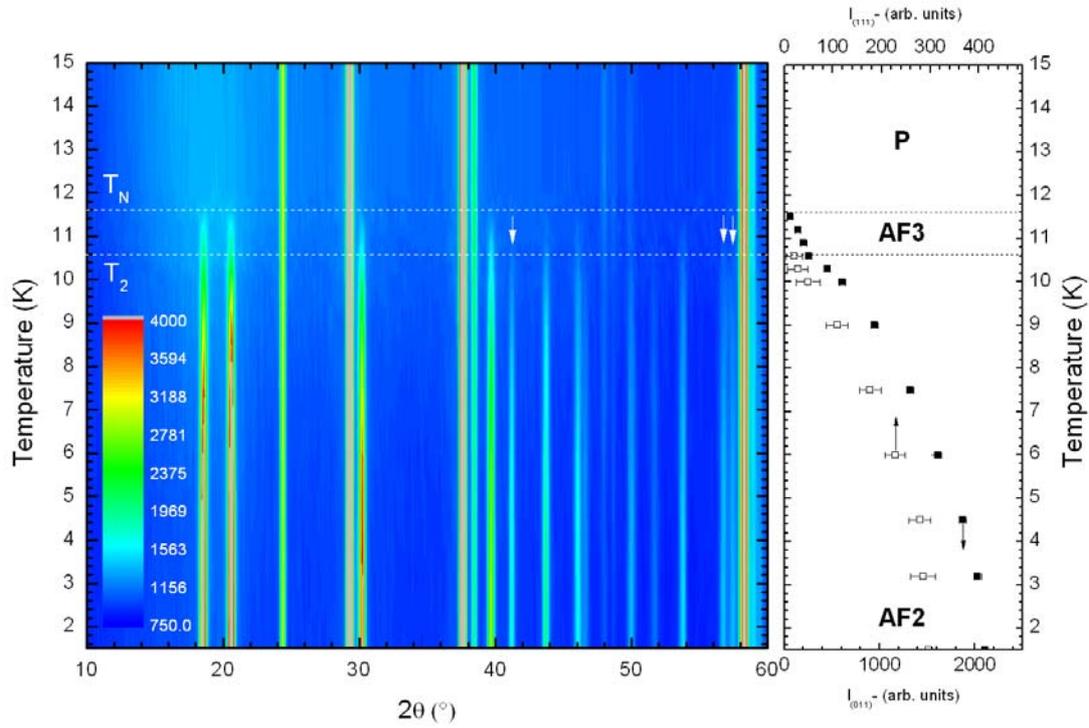

**Figure 15.** Temperature evolution of the G4.1 neutron diffraction patterns (left panel) and of the corresponding integrated intensity of magnetic Bragg peaks (010)⁻ and (111)⁻ (right panel) for $Mn_{0.85}Zn_{0.15}WO_4$. (hkl)⁻ indexation stands for (hkl)-**k**, with k ≈ (-0.209, 0.5, 0.453). The arrows in the left panel indicate the extra magnetic reflections that appear below the AF3-to-AF2 phase transition temperature $T_2(x = 0.15) \approx 10.6$ K.

As shown in Figure 15, additional magnetic peaks develop below $T_2(x = 0.15) \approx 10.6$ K, in agreement with the bulk magnetic measurements. The patterns collected below this temperature are very similar to those previously obtained for $MnWO_4$ in the AF2 phase between $T_2$ and $T_1$.[12] Below $T_2(x = 0.15)$, magnetic peaks can be indexed with a new propagation vector **k** = (-0.209(2), 0.5, 0.453(1)), comparable to the propagation vector **k** = (-0.214, 0.5, 0.457) obtained in the case of $MnWO_4$.[12] The corresponding incommensurate magnetic structure at 1.5 K has been determined by Rietveld refinement using symmetry adapted modes derived from representation analysis performed with the BasIreps program.[19] Table 1 lists the characters of the two one-dimensional irreducible representations of the little group $G_k$. The magnetic representation $\Gamma_m$ calculated for the Wyckoff 2*f* position of the Mn atom in the *P*2/*c* space group contains three times each representation, so that there are three basis functions for each representation. Table 2 lists the corresponding basis vectors. As was already pointed out by Lautenschlager et al.[12] in the case of MnWO$_4$, irreducible representation $\Gamma_2$ (which describes the magnetic structure AF3 existing for $T_2 < T < T_N$ in MnWO$_4$ and Mn$_{0.85}$M$_{0.15}$WO$_4$, M = Zn, Mg) fails to provide a good agreement with the experimental data. In order to obtain a satisfying result, it is necessary to use a linear combination of the two irreducible representations $\Gamma_1$ and $\Gamma_2$. There are accordingly two indistinguishable moment configurations, with identical structure factors, that give a very good agreement factor $R_{mag}$ = 3.37% with the data (Figure 16). Both correspond to a coupling of the *x* and *z* magnetic components following $\Gamma_2$, and of the *y* components following $\Gamma_1$. The first model is obtained by mixing $\psi_1$, $\psi_3$ and $\psi_2'$ with real components and corresponds therefore to a sinusoidal structure. The second model, which is assumed to be the correct one according to the magnetic properties and dielectric properties of Mn$_{0.85}$M$_{0.15}$WO$_4$ (M = Zn, Mg) is obtained by mixing $\psi_1$ and $\psi_3$ with real components and $\psi_2'$ with an imaginary component. This describes an incommensurate spiral structure with an elliptical modulation, in which the spin rotation envelope is perpendicular to the (*a*, *c*) plane, with the moments canted with regards to the *a* axis by about 34°. The elliptical parameter *p* is 0.87. The ordered component maximum is close to 4.9 $\mu_B$. These results are similar to those observed of MnWO$_4$.[12] In the incommensurate AF2 and AF3 phases of MnWO$_4$, the magnetic moment $\boldsymbol{m_i}$ on the *i*th site (i = 1 or 2) at $\boldsymbol{R_i}$ is described as $m_i = m_{easy}\cos(2\pi k R_i) + m_b \sin(2\pi k R_i)$, where $m_{easy}$ and $m_b$ are the components parallel to the magnetic easy axis and crystallographic b-axis, respectively. The easy axis lies in the ac plane forming an angle of $\alpha$ = 34° with the a axis. Both the m$_b$ component and the ellipticity $p = m_b/m_{easy}$ are zero in the sinusoidal AF3 phase, whereas m$_b$ and *p* have nonzero values in the spiral AF2 phase. The refined magnetic moment and ellipticity have been reported to be $|m| \approx 4.5$ $\mu_B$ and $p \approx 0.8$ at 9 K, respectively.[12]

**Table 1.** Irreducible representations of the propagation vector **k** = (-0.209(2), 0.5, 0.453(1)) in $P2/c$. The magnetic representation $\Gamma_m$ contains three times each representation, $\Gamma_m = 3\Gamma_1 \oplus 3\Gamma_2$. $a = \exp(-i\pi k_z)$.

|            | 1 | $c\ x, 0, z$ |
|------------|---|--------------|
| $\Gamma_1$ | 1 | $a$          |
| $\Gamma_2$ | 1 | $-a$         |

**Table 2.** Basis functions for axial vectors associated with irreducible representations $\Gamma_1$ and $\Gamma_2$ for Wyckoff $2f$ site.

| $\Gamma_1$ | $(x, y, z)$ Mn$_1$ | $(x, -y, z+\frac{1}{2})$ Mn$_2$ | $\Gamma_2$ | $(x, y, z)$ Mn$_1$ | $(x, -y, z+\frac{1}{2})$ Mn$_2$ |
|---|---|---|---|---|---|
| $\psi_1$ | (1 0 0) | ($a^*$ 0 0) | $\psi_1'$ | (1 0 0) | ($-a^*$ 0 0) |
| $\psi_2$ | (0 1 0) | (0 $-a^*$ 0) | $\psi_2'$ | (0 1 0) | (0 $a^*$ 0) |
| $\psi_3$ | (0 0 1) | (0 0 $a^*$) | $\psi_3'$ | (0 0 1) | (0 0 $-a^*$) |

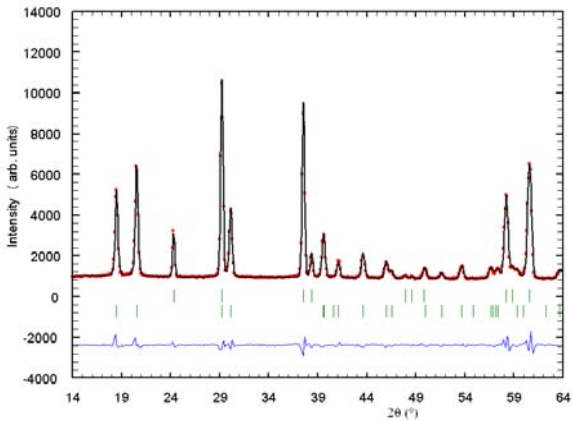

**Figure 16.** Rietveld refinement of the neutron diffraction data of Mn$_{0.85}$Mg$_{0.15}$WO$_4$ at 1.5 K (experimental data : open circles, calculated profile : continuous line, allowed crystal and magnetic Bragg reflections : vertical marks. The difference between the experimental and calculated profiles is displayed at the bottom of the graph).

Overall, this NPD study shows that the commensurate AF1 magnetic structure of MnWO$_4$ is suppressed by the substitution in Mn$_{0.85}$M$_{0.15}$WO$_4$ (M = Zn, Mg) at T = 1.5 K. A spin-spiral phase similar to that observed in the AF2 state of MnWO$_4$ is probably stabilized by the nonmagnetic doping below T$_2$(x = 0.15) ≈ 10.6 K.

## 4. Discussion

Our results described in the previous section show that, in essence, the properties of Mn$_{1-x}$M$_x$WO$_4$ (M = Mg, Zn) are independent of the nature of the nonmagnetic ions Mg$^{2+}$ and Zn$^{2+}$. To explain this finding, we first consider the effects of the Mg and Zn substitutions on the crystal lattice putting aside the effects on the spin lattice. As the substitution contracts the crystal lattice of MnWO$_4$, a low doping could be equivalent to applying an external pressure. According to a recent experimental study on the pressure effect in MnWO$_4$,[25] the "chemical pressure" associated with the substitution would stabilize the non-polar AF1 phase. Although only a slight local structural distortion is expected from the small difference in ionic radii between Mn$^{2+}$ and M$^{2+}$ (M = Mg, Zn), the presence of M$^{2+}$ dopants at the Mn$^{2+}$ sites will introduce chemical disorder. The latter could affect the phase transitions in MnWO$_4$ since a substantial spin-lattice interaction is necessary to explain the ferroelectricity associated with the AF2 spiral magnetic order. The existence of such a spin-lattice coupling has been experimentally proven by a recent synchrotron X-ray diffraction study,[7] which revealed lattice modulations in the ferroelectric AF2 and paraelectric AF3 phases ; the lattice propagation vector of each phase is twice the magnetic propagation vector. Furthermore, high-resolution thermal expansion measurements[25] showed clear anomalies of all three lattice parameters at T$_N$ and T$_1$, and a discontinuous volume change across T$_1$ where a locking of the magnetic modulation with the lattice occurs. Indeed, the first-order AF1-to-AF2 transition at T$_1$ in MnWO$_4$ is more than a mere magnetic transition. In real solids, inevitable imperfections invariably smear first order phase transitions over certain temperature or pressure intervals. When the doping level is intentionally increased, the first-order phase transition can even be suppressed. As a matter of fact, it was observed that the impurities in flux-grown MnWO$_4$ crystals reduce the AF2-to-AF1 transition temperature $T_1$ but don't modify the Néel ($T_N$) and AF3-to-AF2 ($T_2$) phase-transition temperatures.[5]

We now turn to the effects of Mg/Zn doping on the magnetic properties. In essence, the doping of nonmagnetic ions at the $Mn^{2+}$ sites decreases not only the number of spin exchange interactions along the zigzag $MnO_4$ chains along the c-direction but also that between these chains along the a- and b-directions (see below). Nevertheless, the doped compounds remain almost as spin-frustrated as the undoped one, because the frustration parameter $|\theta|/T_N$ varies only slightly (i.e., $|\theta(x)|/T_N \approx 6$ for x = 0, and $|\theta(x)|/T_N \approx 5$ for x = 0.3). All doped samples exhibit a paramagnetic susceptibility behavior consistent with geometric spin frustration, since the mean-field Curie-Weiss regime extends to temperature significantly lower than what could be expected from the absolute Weiss temperature. If both the undoped and doped compounds were not spin-frustrated, the $T_N$ would be of the same order of magnitude as $|\theta|$ for a three-dimensional magnetic system, or the susceptibility curve would exhibit a broad maximum in the paramagnetic regime for low dimensional magnetic systems.

The linear reduction of $T_N$ and $T_2$ shown in Figure 11 is well explained by the prediction of mean-field theory for randomly diluted magnets, $T_c(x) = (1-x)*T_c(0)$, which predicts that the long-range ordering temperature decreases linearly with increasing the amount of non-magnetic dopants until all magnetic ions disappear. For non-frustrated systems, a reduction faster than predicted by the mean-field theory is expected for dopant concentration lower than the percolation threshold.[26] Therefore, the magnetic dilution explains the reduction of the ferroelectric critical temperature, if this transition is driven by magnetic order and is associated with the onset of the AF2 order as in $MnWO_4$.

As in $MnWO_4$, the low-temperature magnetic ordering in the doped systems should arise from competing magnetic interactions in the presence of weak magnetic anisotropy at each $Mn^{2+}$ site.[12,15] The outcome of this competition is likely to be modified due to the suppression of magnetic couplings by the site dilution. Nevertheless, our findings show that the multiferroic state is supported even for a doping concentration as high as x = 0.3. Indeed the absence of dielectric dispersion and the sharpness of the ferroelectric peak, even for the most substituted samples, support a long range ferroelectric order which is connected to a AF2-like spin-spiral magnetic structure. Assuming a random distribution of the nonmagnetic $M^{2+}$ ions in $Mn_{1-x}M_xWO_4$, the corresponding distribution of $MnO_4$ chain segments along the c-axis is characterized by an average segment length $L \approx 1/x$. This means that the multiferroic state exists even when short segments are created by the dilution. And this implies that the nearest-neighbor intra-chain spin exchange interaction is not crucial in stabilizing the multiferroic state. The next-nearest neighbor intrachain and interchain couplings should be important for the multiferroicity of $Mn_{1-x}M_xWO_4$.

It is therefore of interest to discuss the above observation from the viewpoint of the spin exchange paths and their values in $MnWO_4$. The magnetic properties of $MnWO_4$ have been described by nine spin exchange parameters,[15,27] i.e., four exchange interactions ($J_1 - J_4$) within each layer of $Mn^{2+}$ ions parallel to the bc-plane (hereafter, the //bc-layer of $Mn^{2+}$ ions) and five exchange interactions ($J_5 - J_9$) between adjacent //bc-layers of $Mn^{2+}$ ions (Figure 17). Ehrenberg *et al.*[27] interpreted the results of their inelastic neutron scattering measurements for $MnWO_4$ in terms of these nine spin exchange parameters. In their study, $J_1$ is antiferromagnetic (AFM) while $J_2$ is ferromagnetic (FM), so that the spin exchanges within a zigzag chain of $Mn^{2+}$ ions parallel to the c-direction (hereafter, the //c-chain of $Mn^{2+}$ ions) are not spin-frustrated. This is not consistent with the experimental observation that each //c-chain of $Mn^{2+}$ ions exhibits a spiral spin order in the AF2 state, because the noncollinear spin order requires the presence of substantial spin frustration.[28,29] In the AF2 state a spiral spin order occurs along the a-direction as well, implying that the spin exchange interactions between //c-chains of $Mn^{2+}$ ions along the a-direction are also spin-frustrated. The spin exchange parameters extracted on the basis of first principles density functional theory (DFT) calculations[15] show that both $J_1$ and $J_2$ are AFM, and $J_2$ is stronger than $J_1$ in strength. This gives rise to spin frustration within each //c-chain of $Mn^{2+}$ ions. Furthermore, since $J_2$ is substantial, the intrachain spin frustration would not be easily destroyed by introducing a nonmagnetic dopant (e.g., $Mg^{2+}$ or $Zn^{2+}$) into the zigzag chain of $Mn^{2+}$ ions, because the two chain segments of $Mn^{2+}$ ions separated by a diamagnetic dopant $M^{2+}$ can still interact across the dopant via $J_2$. The DFT calculations also show[15] that the interactions between the //c-chains of $Mn^{2+}$ ions along the a-direction are spin frustrated. This also indicates that the interlayer spin frustration along the a-direction is not easily broken by introducing nonmagnetic dopants into the zigzag chain of $Mn^{2+}$ ions. Thus, the persistence of spin frustration along the c- and a-directions under nonmagnetic doping explains why the AF2 state survives the Mg and Zn substitutions.

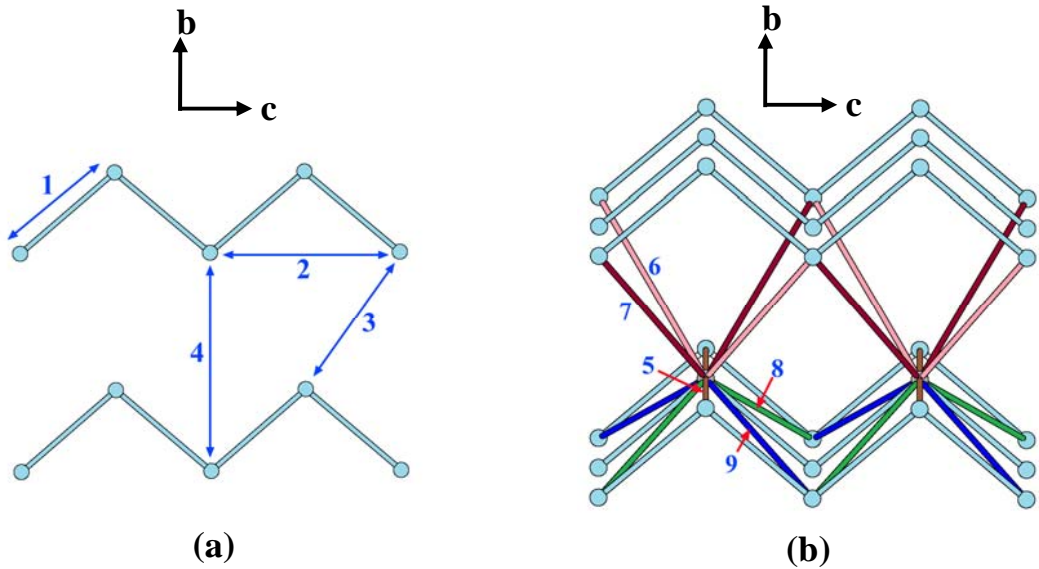

**Figure 17.** (a) Four spin exchange paths $J_1 - J_4$ in $MnWO_4$ within each layer of $Mn^{2+}$ ions parallel to the bc-plane. (b) Five spin exchange paths $J_5 - J_9$ between adjacent //bc-layers of $Mn^{2+}$ ions in $MnWO_4$. The numbers 1 – 9 refer to the spin exchange paths $J_1 - J_9$, respectively.

## 5. Concluding remarks

Our study shows that polycrystalline samples of $Mn_{1-x}MWO_4$ (M = Mg, Zn) solid solutions can be obtained for x up to 0.3. The substitution of the nonmagnetic ions $Mg^{2+}$ and $Zn^{2+}$ for the magnetic ions $Mn^{2+}$ result in very similar effects on the magnetic and dielectric properties of $MnWO_4$. These substitutions destabilizes the non-polar magnetic structure AF1 of $MnWO_4$ but do not suppress the AF3-to-AF2 magnetoelectric phase transition. This indicates that the nonmagnetic dopants destroy neither the three-dimensional nature of magnetic interactions nor the spin frustration within each //c-chain and between //c-chains along the a-direction. Simple nonmagnetic dilution effects explain the reduction of the phase-transition temperatures $T_N$ and $T_2$ by Mg and Zn substitutions. Zero-field as well as field-dependent single-crystal studies are necessary to reveal much more detail on the effects of these nonmagnetic substitutions on the low-temperature magnetic and electric properties.

ACKNOWLEDGMENT. We thank French Centre National de la Recherche Scientifique (CNRS) for providing L.M. with a postdoctoral fellowship. The research at IMN and ICMCB was supported in part by the GDR 3163 Program of the French CNRS. M.-H.W thanks U. S. DOE for the financial support (Grant No. DE-FG02-86ER45259).